\documentclass[%
 aip, apl,
 amsmath,amssymb,
 reprint,%
 twocolumn,
]{revtex4-1}

\usepackage{amsmath,amssymb}
\usepackage{lipsum}
\usepackage{bm}
\usepackage{graphicx}
\usepackage{graphics}
\usepackage{amsmath}
\usepackage{array}
\usepackage[caption=false]{subfig}
\usepackage{xcolor}
\usepackage{subfig}
\usepackage{hyperref}
\usepackage[capitalise]{cleveref}
\usepackage{textcomp}

\usepackage{epstopdf}
\usepackage{cases}
\usepackage{amssymb}
\usepackage{multirow}
\usepackage{siunitx}
\usepackage{cases}
\usepackage{tabularx}
\usepackage[top=1.1in, bottom=1.1in, left=1.0in, right=1.0in]{geometry}
\usepackage[utf8]{inputenc}
\usepackage{tikz}
\usepackage{comment}
\usepackage{lineno,hyperref}
\modulolinenumbers[5]
\usepackage{booktabs}
\usepackage{diagbox}
\usepackage{graphicx}
\usepackage{dcolumn}
\usepackage{bm}
\usepackage{graphicx}
\usepackage{dcolumn}
\usepackage{bm}

\usepackage[utf8]{inputenc}
\usepackage[T1]{fontenc}
\usepackage{mathptmx}
\usepackage{etoolbox}

\makeatletter
\def\@email#1#2{%
 \endgroup
 \patchcmd{\titleblock@produce}
  {\frontmatter@RRAPformat}
  {\frontmatter@RRAPformat{\produce@RRAP{*#1\href{mailto:#2}{#2}}}\frontmatter@RRAPformat}
  {}{}
}%
\makeatother
\begin{document}
\preprint{AIP/APL}

\title{Extended Multi-Temperature Model for Electron--Phonon Coupling and Ultrafast Thermal Transport in Graphene}
\author{Houssem Rezgui}
\email{houssem.rezgui@inl.int}
\affiliation{International Iberian Nanotechnology Laboratory (INL), Braga 4715-330, Portugal}

\author{Chuang Zhang}
\email{zhangc520@hdu.edu.cn}
\affiliation{Department of Physics, School of Sciences, Hangzhou Dianzi University, Hangzhou 310018, China}

\author{Clivia Sotomayor-Torres}
\email{clivia.sotomayor@inl.int}
\affiliation{International Iberian Nanotechnology Laboratory (INL), Braga 4715-330, Portugal}

\date{\today}
%
%

\begin{abstract}

Ultrafast thermal transport in low-dimensional materials challenges traditional diffusive models due to reduced scattering, strong electron-phonon coupling, and pronounced non-equilibrium effects. To address these complexities, we extend the macroscopic multi-temperature model by incorporating non-diffusive and non-local phenomena, treating electrons, optical phonons, and acoustic phonons as coupled but thermally distinct subsystems. We benchmark this enhanced framework against the multi-temperature Boltzmann transport equation, enabling detailed resolution of branch-dependent energy relaxation and identifying bottlenecks in thermalization. This approach provides a more accurate and comprehensive description of heat flow in emerging materials, offering novel insights into phonon dynamics and electron-phonon interactions. These theoretical advances pave the way for the improved design and optimization of next-generation nanoelectronic and photothermal devices.


\end{abstract}

\maketitle

\begin{figure}[htb]
\centering
\includegraphics[width=0.8\columnwidth,clip=true]{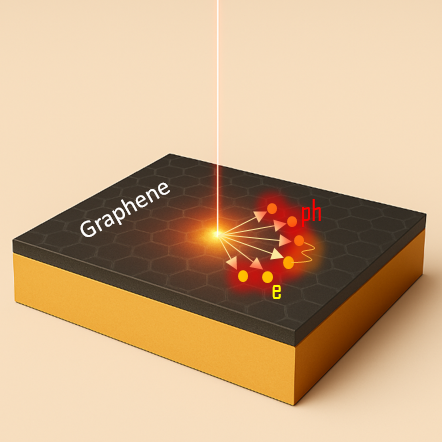}
\caption{Schematic illustration of electron-phonon (e-ph) coupling in photoexcited graphene~\cite{exp_photon_excitation2021,nano_letters_2017_non_equilibrium,Distinguishing_AdvSci_2020,beardo_hydrodynamic_EP_2025,ZHANG2024,Lu2018}. The ultrafast laser pulse rapidly heats electrons, which subsequently transfer energy to optical and acoustic phonon modes.}
\label{fig:figure1}
\vspace{0.5em}
\end{figure}

\noindent\hspace*{1em}Graphene's extraordinary thermal, electrical, and mechanical properties have sparked intense interest in a wide range of scientific and technological domains ~\cite{Balandin2008,Balandin2011,Schwierz2010,Papageorgiou2017}. 
Among these, understanding thermal transport on ultrafast timescales is crucial for applications in high-speed electronics and energy conversion devices ~\cite{pop_energy_2010,Cahill2014}. 
As device dimensions shrink and operational speeds increase, understanding the mechanisms governing heat dissipation at ultrafast timescales becomes increasingly critical.
\noindent\hspace*{1em}A key process underlying thermal relaxation in photoexcited graphene is electron--phonon ($e-ph$) coupling, namely the interaction between energetic charge carriers and lattice vibrations~\cite{exp_photon_excitation2021,nano_letters_2017_non_equilibrium,Distinguishing_AdvSci_2020,beardo_hydrodynamic_EP_2025,ZHANG2024,Lu2018}. A laser pulse rapidly heats electrons, initiating energy transfer to phonons via electron--phonon coupling, as shown in~\cref{fig:figure1}. 
Upon ultrafast laser excitation, electrons in graphene can reach temperatures of several thousand kelvins within femtoseconds. 
The subsequent relaxation of this non-equilibrium state is governed by the transfer of energy from hot electrons to various phonon modes~\cite{Waldecker2016,Laitinen2014,PhysRevLett.130.256901}. In graphene, this process is particularly complex due to its unique phonon spectrum, the presence of strongly coupled optical phonons, and the coexistence of multiple relaxation channels involving acoustic phonons and substrate interactions~\cite{PhysRevLett.113.235502,PhysRevB.97.165416,PhysRevB.93.125432,exp_photon_excitation2021,nano_letters_2017_non_equilibrium,Distinguishing_AdvSci_2020}.
\noindent\hspace*{1em}Traditionally, the two-temperature model (TTM) has been employed to describe ultrafast thermal transport in metals and some semiconductors ~\cite{Anisimov1974,PhysRevLett.59.1460,TTM_EP_AIPX2022,PhysRevLett.119.136602,ZHANG2024}. 
The TTM simplifies the system into two coupled reservoirs: the electrons and the lattice, each characterized by its own temperature and connected by a phenomenological coupling constant ~\cite{Qiu1993,Anisimov1974,PhysRevLett.59.1460,TTM_EP_AIPX2022,PhysRevLett.119.136602}. Although this model has been instrumental in describing heat transfer in bulk systems, it does not capture the subtleties of the ultrafast dynamics of graphene. 
In particular, it overlooks the fact that different phonon branches (e.g., optical and acoustic) can remain out of equilibrium with each other for extended periods, and that energy transfer is neither instantaneous nor uniform across all modes ~\cite{Carpene2006,Waldecker2016,exp_photon_excitation2021,nano_letters_2017_non_equilibrium,Distinguishing_AdvSci_2020}.

\noindent\hspace*{1em}To address the limitations of the traditional TTM, a multi-temperature model (MTM) was developed, introducing separate temperatures for hot electrons, optical phonons, and acoustic phonons to better capture the sequence of ultrafast energy transfer events, including the optical phonon bottleneck and delayed acoustic phonon thermalization~\cite{PhysRevB.93.125432,Lu2018}. 
At the same time, the classical Fourier law of heat conduction, which assumes local and diffusive transport with constant thermal conductivity, becomes inadequate for low-dimensional materials ~\cite{xu_length-dependent_2014,Guo2018,Chen2021}. In systems like suspended and pure graphene, phonons can travel micrometer distances without scattering, resulting in ballistic or quasi-ballistic heat transport and size-dependent thermal conductivity, which shows clear deviations from Fourier behavior ~\cite{PhysRevBLindsay10,doigraphene14,xu_length-dependent_2014}. Additionally, the role of specific phonon modes, such as flexural acoustic phonon modes (ZA), becomes increasingly important in these systems, as their contributions are no longer averaged out as in bulk or diffusive models ~\cite{PhysRevBLindsay10}. These complexities necessitate more advanced theoretical frameworks that can resolve microscopic, non-equilibrium heat transport, particularly under the ultrafast conditions probed by femtosecond pump-probe experiments. In this letter, we provide a more accurate description of non-diffusive thermal transport, particularly in ultrafast phonon dynamics, thereby motivating the need for advanced modeling approaches, such as the multi-temperature framework, beyond the TTM.

Although Fourier's law works well for many macroscopic scenarios, it assumes instantaneous heat propagation with a nonphysical infinite speed ~\cite{RevModPhysJoseph89}. This assumption breaks down in non-equilibrium thermodynamics, particularly in micro- and nanoscale systems or cryogenic environments ~\cite{WangMr15application,PhysRev.148.778,Rezgui2023,PhysRevB.111.035406}.
The Guyer–Krumhansl equation (GKE) is an advanced heat conduction model that extends beyond classical Fourier’s law, offering a more accurate description of thermal transport in systems where non-local effects are important. Originally developed in the 1960s by R.A. Guyer and J.A. Krumhansl, the model accounts for both non-local and memory effects in heat flow~\cite{PhysRev.148.766}. While the original GK equation was derived under the assumption of dominant normal phonon scattering, making it suitable mainly for low-temperature applications, more recent formulations have adapted the model for room-temperature conditions by incorporating resistive phonon scattering based on the small perturbation expansion~\cite{WangMr15application,PhysRevB.103.L140301,PhysRevB.98.104304,beardo_hydrodynamic_EP_2025}. This extension enables the GK equation to be used in a wider variety of materials and conditions, particularly in micro- and nanoscale systems at ambient temperatures. The general form of the GK equation for phonon $p$ and electron $e$ is as follows:
\begin{align}
\tau_{p,i} \frac{\partial \mathbf{q}_{p,i}}{\partial t} + \mathbf{q}_{p,i} + \kappa_{p,i} \nabla T_{p,i} &= \ell_{p,i}^2 \left( \nabla^2 \mathbf{q}_{p,i} + 2\nabla(\nabla \cdot \mathbf{q}_{p,i}) \right) \notag \\
\tau_e \frac{\partial \mathbf{q}_e}{\partial t} + \mathbf{q}_e + \kappa_e \nabla T_e &= \ell_e^2 \left( \nabla^2 \mathbf{q}_e + 2\nabla(\nabla \cdot \mathbf{q}_e) \right)
\end{align}
where \( q_{p} \) and \( q_{e} \) are the heat fluxes of phonons and hot electrons, the subscript $i$ is the index of the phonon branch,
\( \tau_{p} \) and \( \tau_{e} \) are the relaxation times, \( \kappa_{p} \) and \( \kappa_{e} \) are the bulk thermal conductivities, \( T_{p} \) and \( T_{e} \) are the temperatures, and \( \ell_p \) and \( \ell_e \) are the non-local mean free paths (MFP). This equation captures both temporal and spatial nonlocal effects and serves as a foundation for constructing a multi-temperature model (MTM) description of ultrafast thermal transport in low-dimensional materials. This equation generalizes Fourier's law by introducing finite propagation speed and spatial effects into the heat flux. The right-hand side accounts for spatial dispersion, which becomes prominent in systems where the mean free path of phonons is comparable to or larger than the system size, as is the case in graphene. In the MTM, the temporal evolution is expressed as
\begin{equation}
\begin{aligned}
C_{e} \frac{\partial T_{e}}{\partial t} &= 
- \nabla \cdot \vec{q}_{e} 
- \sum_{i} G_{ep,i} (T_{e} - T_{p,i}), \\
C_{p,i} \frac{\partial T_{p,i}}{\partial t} &= 
- \nabla \cdot \vec{q}_{p,i} 
+ \ G_{ep,i} (T_{e} - T_{p,i}) \\
&\quad + G_{pp,i} (T_{lat} - T_{p,i}).
\end{aligned}
\end{equation}

where \( C_{p} \) and \( C_{e} \) are the heat capacities of phonons and electrons, respectively, and \( G_{ep,i} \) is the phonon–electron coupling factor.
Combining multi-temperature models with GKE provides a powerful framework for capturing both the energy partitioning among multiple interacting subsystems and the detailed nonequilibrium transport dynamics in ultrafast processes. In this approach, the system is divided into several thermal reservoirs: electrons, optical phonons, and acoustic phonons, each characterized by their own temperatures, as in conventional MTMs. However, rather than assuming instantaneous internal thermalization or relying solely on phenomenological coupling constants, the GKE is employed to describe the time- and space-resolved evolution of heat flux within and between these subsystems. This allows for the inclusion of non-Fourier effects, such as ballistic transport and spatial nonlocality, while preserving the intuitive structure of MTMs. The MTM-GKE can be derived by combining equations (1) and (2), as follows:
\begin{multline}
\tau_e \frac{\partial^2 T_e}{\partial t^2} + \frac{\partial T_e}{\partial t} \\
= \frac{\kappa_e}{C_e}\nabla^2 T_e + 3\ell_e^2 \nabla^2\left(\frac{\partial T_e}{\partial t}\right)
- \sum_{i} \frac{G_{ep,i}}{C_e}(T_e - T_{p,i})
\end{multline}
 \begin{multline}
\tau_{p,i} \frac{\partial^2 T_{p,i}}{\partial t^2} + \frac{\partial T_{p,i}}{\partial t}
= \frac{\kappa_{p,i}}{C_{p,i}}\nabla^2 T_{p,i} + 3\ell_{p,i}^2 \nabla^2\left(\frac{\partial T_{p,i}}{\partial t}\right) \\+ \frac{G_{ep,i}}{C_{p,i}}(T_e - T_{p,i}) + \frac{G_{pp,i}}{C_{p,i}}(T_{lat} - T_{p,i})
\end{multline}

Note that here we define the phonon-electron coupling factor as \( G_{ep,i} = G_{ep,i}^{n} + G_{ep,i}^{SC} \), which represents the total energy transfer between electrons and the phonon branch $i$. Energy exchange in graphene occurs primarily through normal scattering, \( G_{ep}^{n} \), and supercollision scattering, \( G_{ep}^{SC} \). Supercollision scattering involves disorder- or impurity-assisted processes that relax momentum conservation constraints, enabling electrons to emit or absorb phonons over a broader range of momenta. This mechanism enhances electron cooling by offering more efficient energy relaxation pathways compared to momentum-conserving normal scattering. The resulting eMTM-GKE framework offers a balanced approach that combines computational efficiency with the physical accuracy needed to describe ultrafast dynamics in complex materials, particularly in regimes involving strong electron-phonon non-equilibrium states. 

To support the eMTM-GKE and achieve a more accurate description, we also develop a multi-temperature BTE model for electron-–phonon coupling in graphene.~\cite{zhang_MTM_2025},
\begin{align}
\frac{ \partial u_e }{\partial t} + \bm{v}_e \cdot \nabla u_e = \frac{ u_e^{eq}  -u_e }{\tau_e}  - \sum_{i}  G_{ep,i} (T_e - T_{p,i}) ,  \label{eq:epBTE1} \\
\frac{ \partial u_{p,i} }{\partial t} + \bm{v}_{p,i} \cdot \nabla u_{p,i} = \frac{ u_{p,i}^{eq}  -u_{p,i} }{\tau_{p,i} } +  G_{ep,i} (T_e - T_{p,i})  ,
\label{eq:epBTE2}  
\end{align}
where the subscripts $e$ and $p$ represent the electron and phonon, respectively.
$u$ is the distribution function of the energy density; $\bm{v}$ is the group velocity; $\tau$ is the relaxation time; $G_{ep}$ is the electron-phonon coupling constant.
$u_e^{eq}=  C_e T_e/ 2 \pi $ and $u_{p,i}^{eq}= C_{p,i} T_{L}/ 2 \pi $ are the equilibrium states, where $C$ is the specific heat and $T_L$ is the lattice temperature.
Energy is conserved during the electron-electron or phonon-phonon scattering processes.
\begin{align}
\int \frac{ u_e^{eq}  -u_e }{\tau_e }  d\Omega &= 0,  \label{conservation1} \\
\sum_{i }^{} \left( \int \frac{ u_{p,i}^{eq}  -u_{p,i} }{\tau_{p,i} }  d\Omega  \right) &= 0.  \label{conservation2}
\end{align}
where $d\Omega$ represents the integral over the entire solid angle space.
In the diffusive limit, the multi-temperature model could recover the macroscopic multi-temperature model according to the Chapman-Enskog expansion. 
A discrete unified gas kinetic scheme (DUGKS) is used to solve the above multi-temperature BTE~\cite{ZHANG2024,zhang_MTM_2025}. 
Detailed numerical procedures can be consulted in previous papers~\cite{ZHANG2024,zhang_MTM_2025}. 

In this work, both the multi-temperature BTE and MTM-GKE are used to study the ultrafast thermal transport process in single-layer graphene (SLG) materials~\cite{Lu2018,PhysRevB.93.125432}.
At the initial moment, the temperature of SLG materials is $T_0=297$ K.
The system size of SLG in the $x$ and $y$ directions is $L_x$ and $L_y$, respectively.
As shown in Figure 2(a), a Gaussian laser is suddenly loaded onto the central region of the upper boundary, resulting in the boundary temperature of the laser-heated area changing over time.
\begin{align}
T(x,y,t)= T_0 + \Delta T  \exp \left(- \frac{(t-t_{pump})^2}{2 t_s^2 } \right),  \notag  \\
x\in[ -d_{pump}/2, d_{pump}/2], y= 0 ,
\end{align}
where $d_{pump}$ and $t_{pump}$ are the diameter and heating time of the laser, $t_s$ is the full width at half maximum duration of the laser pulse, and $\Delta T$ is the maximum temperature rise.
For laser-unheated areas, the boundary temperature is $T_0$.
All four geometric boundaries of graphene are treated with thermalizing boundary conditions. 
In this study, we set $\Delta T=1000$ K, $t_{pump} =200$ fs, $t_s =85$ fs, $d_{pump}=30$ nm, $L_x=200$ nm, $L_y=50$ nm.
The thermal properties and electron–phonon ($e-p$) coupling parameters derived from first-principles calculations are adopted from \cite{Lu2018,beardo_hydrodynamic_EP_2025} and summarized in Table~\ref{tab:thermal_properties}.

\begin{table}[t]  
\centering
\caption{Thermal Properties and Coupling Factors of Hot Electrons and Different Phonon Branches~\cite{beardo_hydrodynamic_EP_2025,Lu2018}}
\label{tab:thermal_properties}
\scriptsize  
\begin{tabular}{|l|c|c|c|c|c|c|c|}
\hline
\textbf{Properties} & \textbf{Electron} & \textbf{LA} & \textbf{TA} & \textbf{ZA} & \textbf{LO} & \textbf{TO} & \textbf{ZO} \\
\hline
$\kappa$ (W/(m$\cdot$K)) & 50 & 863.0 & 237.9 & 2780.0 & 10.0 & 10.0 & 20.9 \\
\hline
$C$ (MJ/(m$^3 \cdot$K)) & 3.6E-4 & 0.19 & 0.32 & 0.61 & 0.03 & 0.02 & 0.16 \\
\hline
$G_{ep}^{n}$ (TW/(m$^3 \cdot$K)) & -- & 100 & 1 & 0 & 600 & 2700 & 0 \\
\hline
$G_{ep}^{SC}$ (MW/(m$^3 \cdot$K)) & -- & 93.9 & 413.6 & 0 & 0 & 0 & 0 \\
\hline
$G_{pp}$ (TW/(m$^3 \cdot$K)) & -- & 2700 & 1.3E4 & 1900 & 2700 & 1400 & 400 \\
\hline
$\tau$ (ps) & 0.28 & 70.8 & 24.7 & 317 & 10 & 12 & 388 \\
\hline
$\ell$ ($\mu$m) & 0.28 & 1.42 & 0.494 & 6.34 & 0.2 & 0.24 & 7.76 \\
\hline
\end{tabular}
\end{table}

\begin{figure*}[htb]
    \centering
    \subfloat[]{\includegraphics[scale=0.5,clip=true]{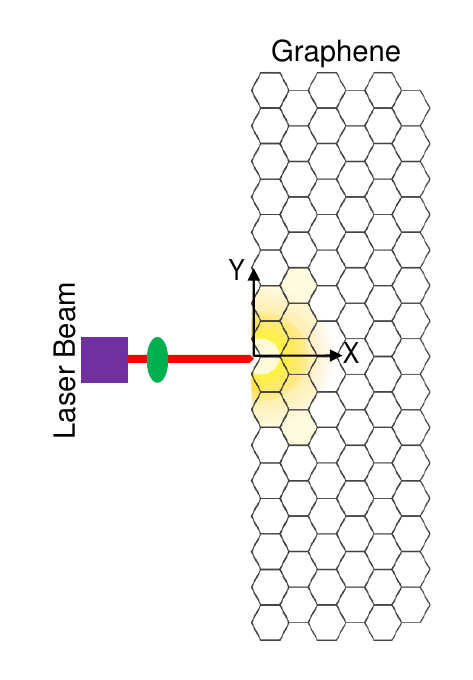}}
    \subfloat[]{\includegraphics[scale=0.48,clip=true]{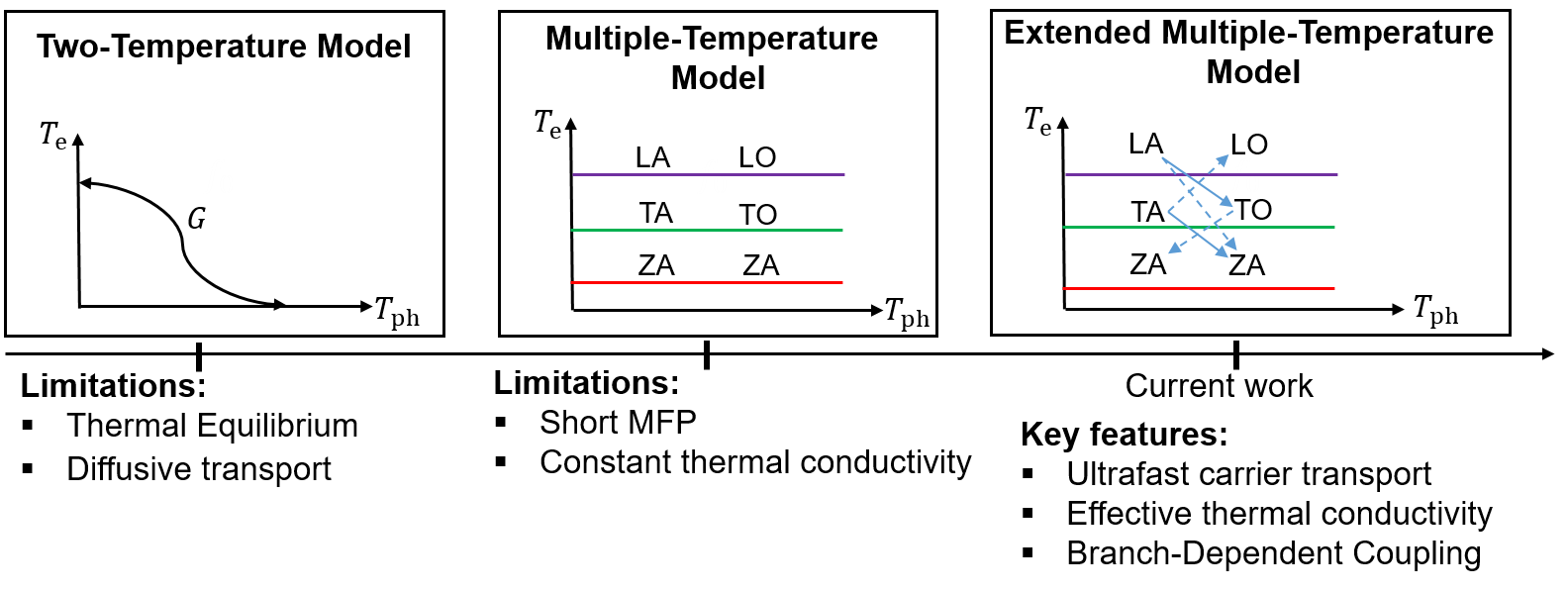}} \\
    \subfloat[]{\includegraphics[scale=0.44,clip=true]{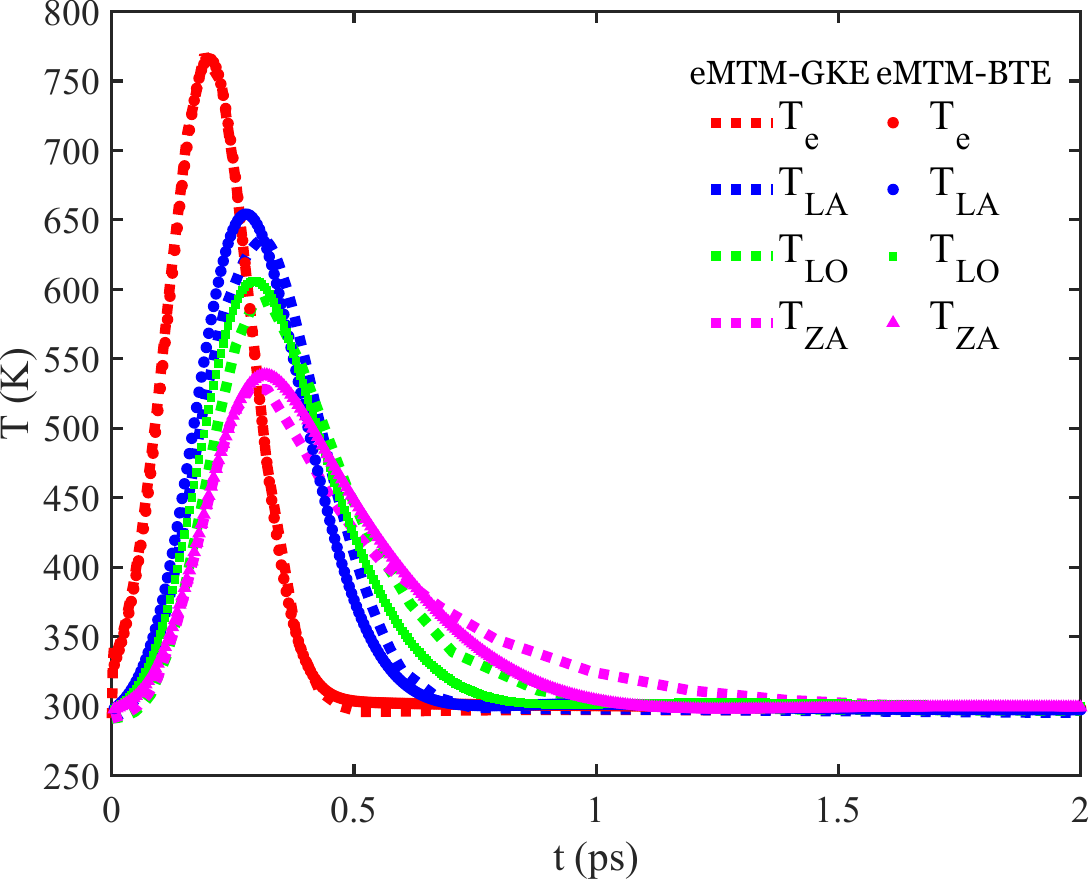}}~~~
    \subfloat[]{\includegraphics[scale=0.44,clip=true]{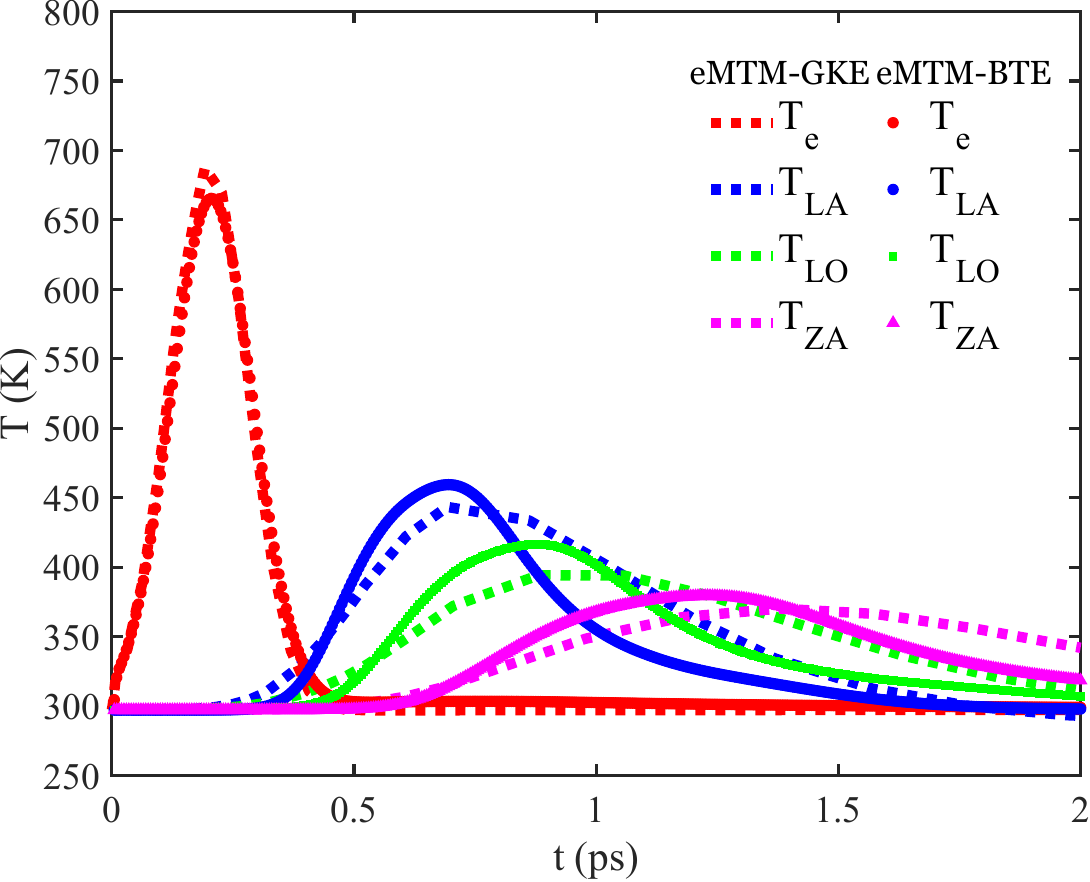}} 
    \caption{(a) Hot-spot generation in single-layer graphene induced by ultrafast laser pulse. (b) Evolutionary history of electron–phonon coupling heat-conduction models, emphasizing the progression from the standard TTM to modern extended MTM formulations.. Temporal evolution of the temperatures at spatial positions (c) $y = 1$~nm and (d) $y = 5$~nm, respectively, from the center of the ultrafast laser excitation.}
    \label{fig:myfigure}
\end{figure*}

Recent advances in theoretical modeling of ultrafast thermal transport in low-dimensional materials, such as graphene, have highlighted the inadequacies of traditional frameworks~\cite{Lu2018,PhysRevB.93.125432}. As illustrated in Figure 2(b), the interaction of a femtosecond laser pulse with graphene initiates non-equilibrium energy transfer processes between electrons and the lattice. The classical TTM simplifies the system into two coupled thermal reservoirs—electrons (with temperature $T_{e}$) and phonons (with temperature $T_{ph}$) linked by a coupling constant $G$. However, this model assumes rapid internal thermal equilibrium within each reservoir and diffusive transport, assumptions that break down under ultrafast excitation and nanoscale confinement. To overcome these challenges, the MTM extends the TTM by treating different phonon branches, both acoustic and optical, as thermally distinct subsystems. Each branch is assigned its own temperature and coupling strength to electrons, enabling the model to capture the complex non-equilibrium dynamics and varying relaxation behaviors observed in time-resolved experiments. The extended Multiple-Temperature Model (eMTM) further enhances this framework by incorporating non-diffusive transport effects, such as ballistic and quasi-ballistic phonon propagation and mode-dependent mean free paths. Unlike the MTM, the eMTM accounts for spatial non-locality and effective thermal conductivity derived from ultrafast carrier and phonon dynamics~\cite{de_tomas_kinetic_2014}. This enables accurate modeling in regimes where local thermal equilibrium is violated, and Fourier’s law fails to describe heat transport.

Figures 2(c) and 2(d) present the temporal evolution of the temperatures of electrons ($T_e$), longitudinal acoustic phonons $T_{LA}$, longitudinal optical phonons $T_{LO}$, and out-of-plane acoustic phonons $T_{ZA}$ at spatial positions $y = 1$~nm and $y = 5$~nm, respectively, from the center of the ultrafast laser excitation. The results were obtained using both the eMTM-BTE and the eMTM-GKE frameworks. Immediately after the femtosecond laser pulse, $T_e$ rises sharply within the first 0.2 ps, reaching a remarkable peak consistent with rapid photoexcitation. Among the phonon branches, the $LA$ phonons exhibit the fastest temperature rise, highlighting their strong coupling to hot electrons and their role as an efficient channel for initial energy dissipation. In contrast, $LO$ phonons show a more moderate temperature increase, while the $ZA$ phonons respond the slowest, reflecting their weaker coupling and longer characteristic relaxation times.
Despite their distinct physical formulations, the eMTM-BTE and eMTM-GKE models exhibit excellent agreement across both spatial positions ($y = 1$~nm and $y = 5$~nm), accurately capturing the transient peak values and relaxation behaviors of all subsystems. This strong consistency confirms the eMTM-GKE as a computationally efficient yet physically reliable alternative to full BTE simulations for modeling ultrafast thermal transport. The pronounced separation in the temperature evolution of different phonon branches, such as the rapid response of $LA$ phonons and the delayed thermalization of $ZA$ modes, highlights the limitations of the classical TTM, which assumes instantaneous equilibrium between electrons and a single lattice temperature. Additional details are provided in the supplementary material (see Figures S1 and S2). These results emphasize the necessity of a multi-temperature framework to capture the inherently non-equilibrium and branch-resolved dynamics of low-dimensional materials like graphene, where spatial non-locality and phonon-mode specificity govern energy dissipation.

\begin{figure}[htb]
\centering
\includegraphics[width=\linewidth]{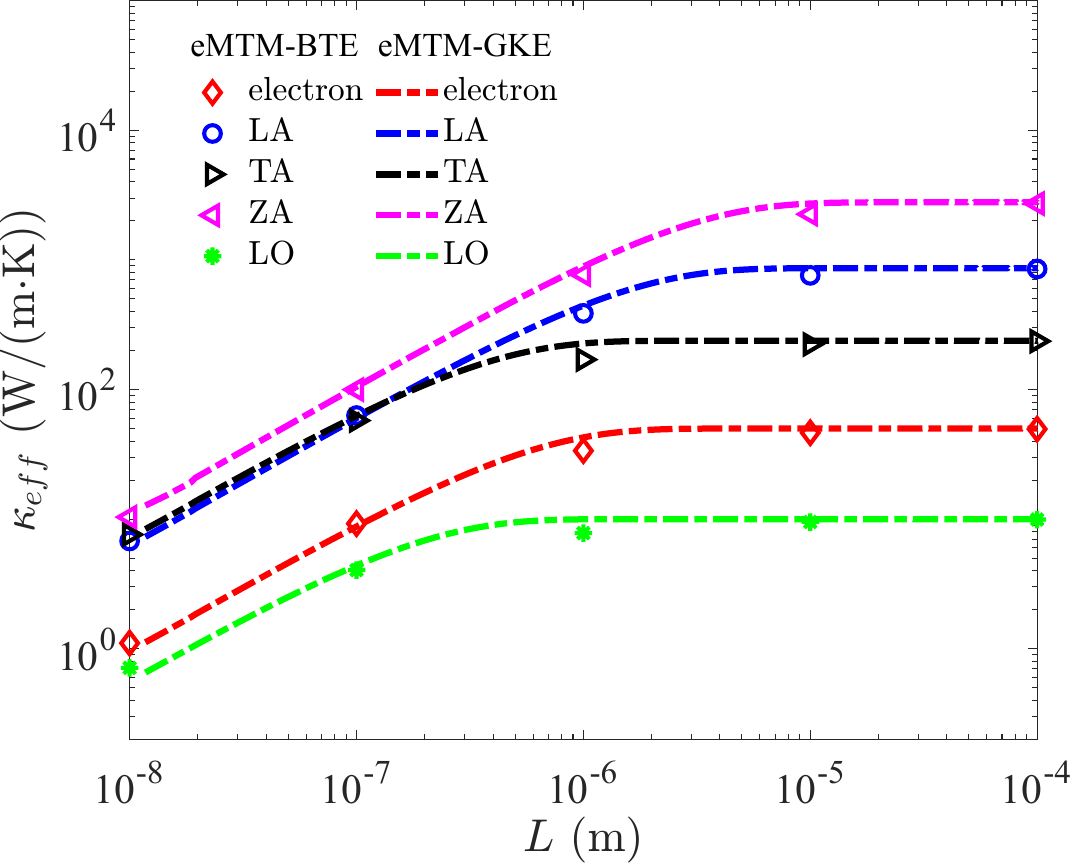}
\caption{Effective thermal conductivity the electron and different phonon branches.}
\label{fig:figure2}
\end{figure}

Figure 3 demonstrates excellent agreement between the eMTM-GKE and eMTM-BTE, highlighting the reliability and predictive power of the proposed solver. Both approaches capture the size effect behavior of the effective thermal conductivity $k_{eff}$ for different energy carriers—electrons, $LA$, $TA$, $ZA$, and $LO$ phonon branches. Notably, the solver accurately reproduces the ballistic-to-diffusive transition, a hallmark of nonlocal transport in low-dimensional systems like graphene. This capability is particularly important for nanoscale thermal transport, where traditional diffusive models fail. The close match between eMTM-GKE and the more computationally intensive eMTM-BTE validates the use of the former as a fast yet physically consistent tool for modeling ultrafast thermal transport, offering an ideal compromise between accuracy and efficiency in predicting branch-resolved thermal conductivity across scales. The corresponding equations used to compute $k_{eff}$ are provided in Section S1 of the Supplementary Material.

\begin{figure}[!htbp]
\centering
\subfloat[$T_{e}$ at 250 fs]{\includegraphics[scale=0.33,viewport=120 10 390 550,clip=true]{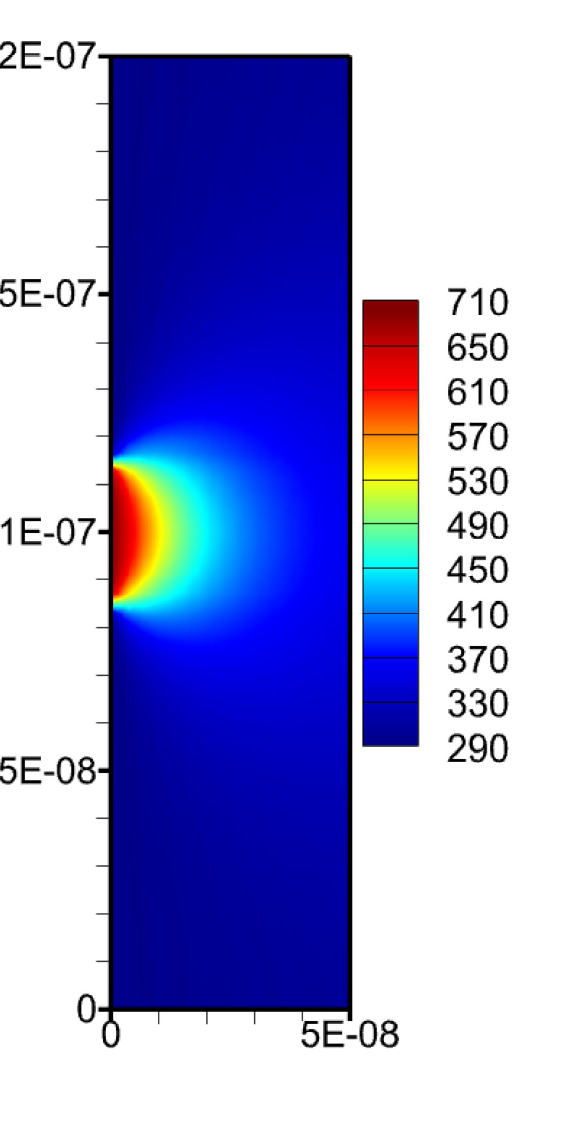}}~~
\subfloat[$T_{e}$ at 250 fs]{\includegraphics[scale=0.33,viewport=120 10 390 550,clip=true]{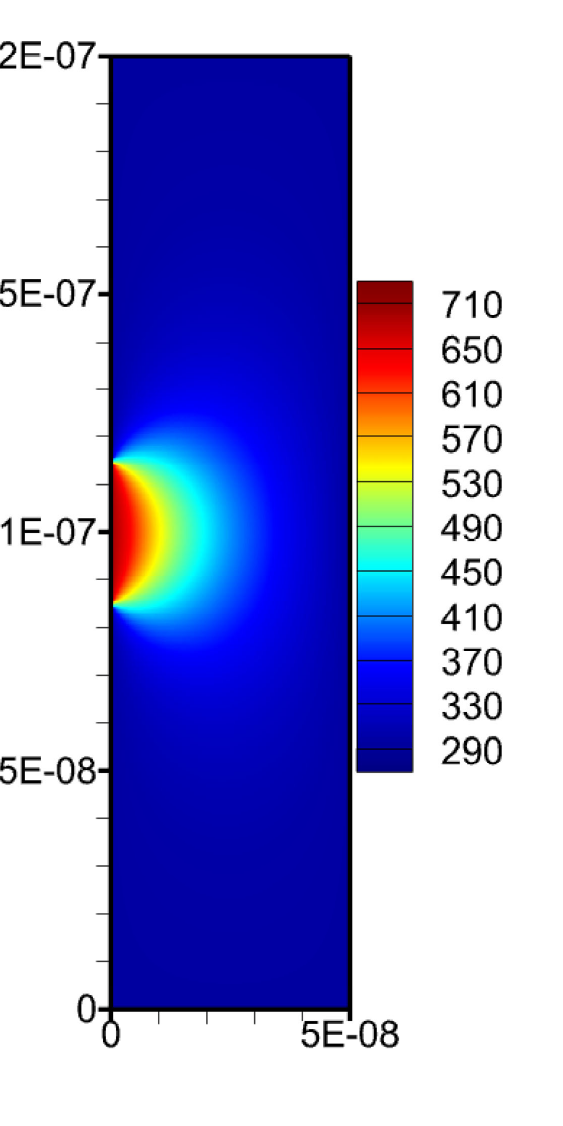}}  \\
\subfloat[BTE $T_{LA}$ at 650 fs]{\includegraphics[scale=0.33,viewport=120 10 390 550,clip=true]{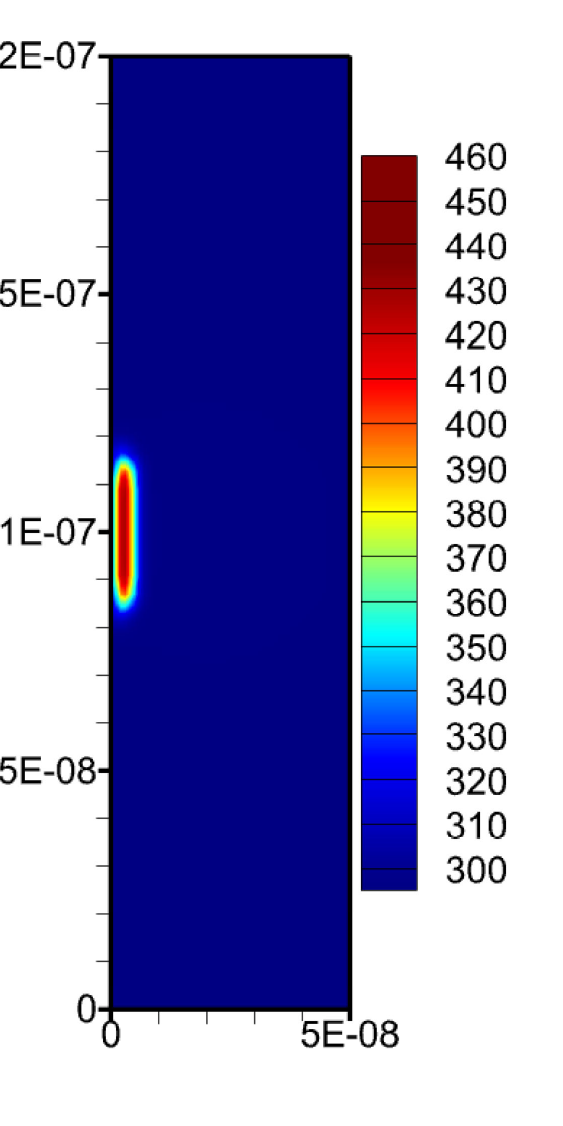}}~~
\subfloat[GKE $T_{LA}$ at 650 fs]{\includegraphics[scale=0.33,viewport=120 10 390 550,clip=true]{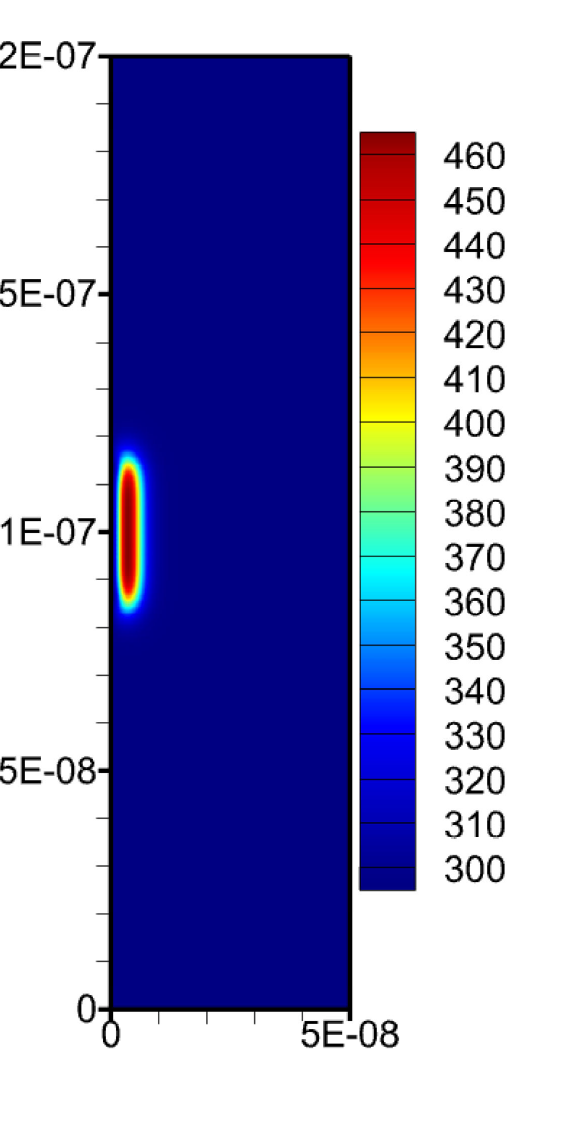}}  \\
\subfloat[BTE $T_{LO}$ at 1000 fs]{\includegraphics[scale=0.33,viewport=120 10 390 550,clip=true]{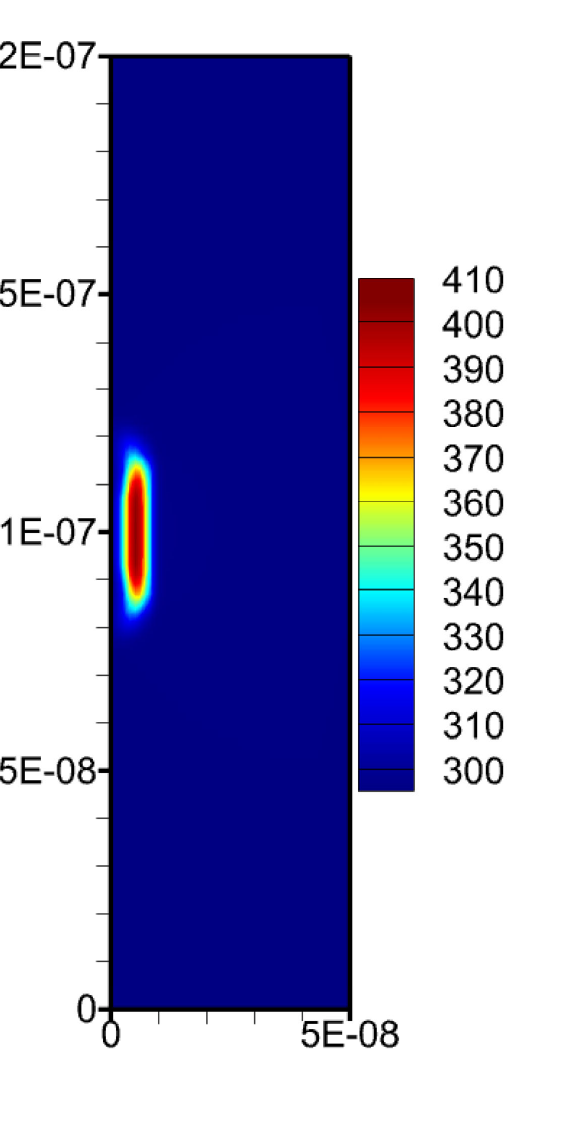}}~~
\subfloat[GKE $T_{LO}$ at 1000 fs]{\includegraphics[scale=0.33,viewport=120 10 390 550,clip=true]{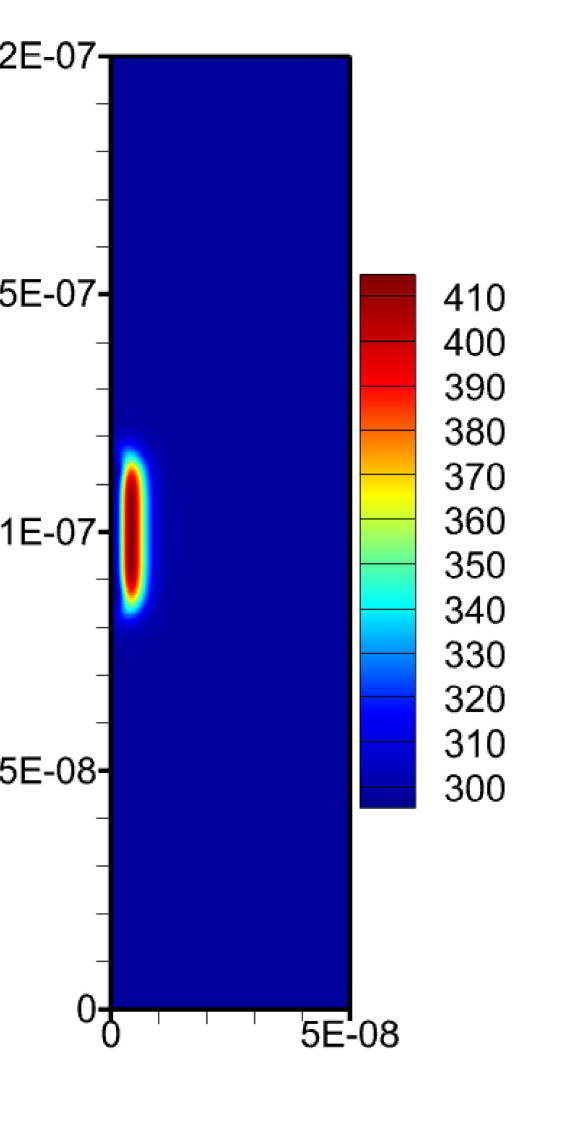}}  \\
\caption{Temperature contours originating from the hot spot. Figures 4(a), (c), and (e) are obtained using the eMTM-BTE, while Figures 4(b), (d), and (f) are obtained using the eMTM-GKE.}
\label{fig:figure3}
\end{figure}

Figure 4 presents temperature contour maps comparing the spatial distributions of $T_e$, $T_{LA}$, and  $T_{LO}$ obtained from the eMTM-BTE and eMTM-GKE approaches following laser excitation. Figure 4(a) shows $T_e$ at 250 fs, where both models demonstrate excellent spatial agreement. They accurately capture hot-spot formation and subsequent heat propagation, with very similar temperature gradients and decay profiles, confirming both models' ability to describe ultrafast electron transport dynamics. Figure 4(b) shows $T_{LA}$ at 650 fs, again revealing strong consistency between the two methods. Both the eMTM-BTE and eMTM-GKE models capture localized heating patterns with closely matching temperature distributions. Similarly, Figure 4(c) shows $T_{LO}$ at 1000 fs, with both approaches reproducing the confined energy localization near the excitation center and closely aligned temperature profiles. The elevated $T_{LO}$ levels, particularly near the excitation region, reflect the preferential energy transfer from hot electrons to high-frequency optical phonon modes due to strong electron–-phonon coupling. This spatial localization illustrates the “hot-phonon bottleneck” effect commonly observed in graphene, where optical phonons accumulate energy more rapidly than this energy can be dissipated to acoustic modes. The energy redistribution follows a sequential pathway. First, electrons absorb energy from the laser pulse and rapidly thermalize. Then, energy is transferred to optical phonons via strong coupling. Finally, acoustic phonons are heated through anharmonic scattering, allowing heat to dissipate across the lattice. This distinct temporal sequence of rapid electron heating within hundreds of femtoseconds, followed by slower phonon heating on the picosecond scale, provides clear evidence of nonequilibrium energy exchange. The ability of both models to resolve this decoupled thermalization pathway underscores their strength in capturing ultrafast energy flow across subsystems and validates their reliability for modeling branch-resolved thermal transport in graphene. Additional details are provided in Figure S3 of the Supplementary Material, which illustrates branch-resolved thermal transport modeling and highlights the main characteristics of the current methods.

In this work, we have demonstrated that both the eMTM-BTE and eMTM-GKE solvers are accurate and efficient frameworks to model ultrafast thermal transport in graphene. The two approaches show excellent agreement in predicting size-dependent effective thermal conductivity and branch-resolved electron and phonon temperature dynamics. Importantly, they are capable of capturing the transient energy redistribution pathways resulting from electron–-phonon coupling. This includes the sequential transfer of energy from hot electrons to optical phonons and subsequently to acoustic modes. This enables detailed time- and mode-resolved mapping of nonequilibrium thermal processes, which are essential to understand heat dissipation in low-dimensional systems. These results provide a strong foundation for future work involving experimental validation and the application of both models to advanced nanoscale thermal management.

\section*{Acknowledgements}

C. Z. acknowledges the support of the National Natural Science Foundation of China (52506078).
The authors acknowledge Beijing PARATERA Tech CO., Ltd. for the HPC resources.

\section*{Supplementary Material}
The Supplementary Material compares the two core theoretical frameworks introduced in the main manuscript, eMTM-BTE and eMTM-GKE. These models describe ultrafast electron–phonon dynamics and non-diffusive heat conduction in graphene, each offering distinct computational and theoretical trade-offs.
\section*{References}
\bibliographystyle{elsarticle-num-names}
\bibliography{REF}
\end{document}